\documentclass{article}
\usepackage{epsf,latexsym}

\begin{document}
\twocolumn[
\begin{flushright}
        Alberta-THY 22-96\\
    gr-qc/9608034
\end{flushright}

\begin{center}
        {\Large \bf Numerical investigation of black hole
interiors}\vspace{1cm}\\
        S. Droz{\footnotemark[2]}
        \vspace{1cm}\\
        Theoretical Physics Institute, University of  Alberta,\\
        Edmonton, Alberta, Canada T6G 2J1
\end{center}
\vspace{1cm}
\begin{abstract}
Gravitational perturbations which are present in any realistic
stellar
collapse to a black hole, die off in the exterior of the hole,
but
experience an infinite blueshift in the interior. This is believed to
lead to a slowly contracting lightlike scalar curvature singularity,
characterized by a divergence of the hole's (quasi-local) mass
function
along the inner horizon.

The region near the inner horizon is described to great accuracy by
a plane wave spacetime. While Einstein's equations for this metric
are still too complicated to be solved in closed form it is
relatively
simple to integrate them numerically.

We find for generic regular initial data  the predicted mass
inflation type
null singularity, rather than a spacelike singularity. It thus seems
that mass inflation indeed represents a generic self-consistent
picture
of the black hole interior.
\end{abstract}
\vspace{1cm}
] 
\section{Introduction}
\footnotetext[2]{Address after September 1, 1996:\\Department of
Physics, University of Guelph\\Guelph, Ontario, Guelph N1G 2W1}

 The strong cosmic censorship
hypothesis suggests the existence of a central spacelike curvature
singularity at the core of the black hole. The Kerr solution
which describes the exterior of the hole very accurately however
possesses a timelike singularity in its interior \cite{carter:66}.
This singularity lies behind  an interior Cauchy horizon,
which in the Kerr case coincides with the inner horizon of the black
hole. Penrose argued on rather general grounds that this inner
horizon should be unstable \cite{penrose:68}.
Perturbations, present in any realistic stellar collapse
\cite{price:72}, falling into the black hole experience an infinite
blueshift as they near the inner horizon. At the same time scattered
radiation crossing the Cauchy horizon cause its contraction. These
two effects lead to the formation of a lightlike scalar curvature
singularity along the Cauchy horizon \cite{poisson:89,bonanno:95}
which slowly tapers on to the final spacelike central singularity.
While the latter is believed to be of the oscillatory BKL type
\cite{belinski:70} the former is rather weak and smooth in the sense
that
is fully characterized by the divergence of a single function along
the Cauchy horizon, which can be identified with the (quasi-local)
mass function of the black hole. Hence one speaks also of a {\em mass
inflation} singularity.

Mass inflation is well established in the spherical case, both
analytically \cite{poisson:89,bonanno:95} and numerically
\cite{brady:95:2}.
The analysis of the generic, non-spherical case on the other hand is
much more difficult. Earlier attempts were either somewhat
inconclusive
\cite{ori:92} or assumed additional properties like slow rotation
\cite{bonnano:95}. Nevertheless these analyses all seem to
agree with  the spherical mass inflation picture.

Only recently has the full non-spherical problem been tackled
\cite{brady:96:2}.  In this latest approach the full set of
Einstein's
equations are expanded in an inverse curvature series,
utilizing a recently developed double null formalism \cite{brady:96}.

However the equations governing the shear (which encodes the
gravitational perturbations leading to mass inflation) are, not
surprisingly, a pair of nonlinear coupled partial differential
equations, even
to lowest order in the above mentioned expansion. This makes the
analysis of the structure of the Cauchy horizon singularity
exceedingly
difficult. One way to handle this problem is a numerical
analysis of the equations.

The purpose of this paper is to report the results of such a
numerical
experiment. The numerical solution of the Einstein's equations,
supplied with the correct boundary conditions indeed confirms the
analytic prediction and thus places the mass inflation scenario on
firm
ground.

\section{The plane wave model}

The investigation of the Cauchy horizon is simplified by two facts.
First, descent into a black hole is necessarily progress in time.
Remember that the radial coordinate $r$ in the Kerr
case becomes timelike in the interior. Thus we don't need to know the
(quantum)
physics of  central core to analyze the outer layers of the black
hole interior.

\epsffile{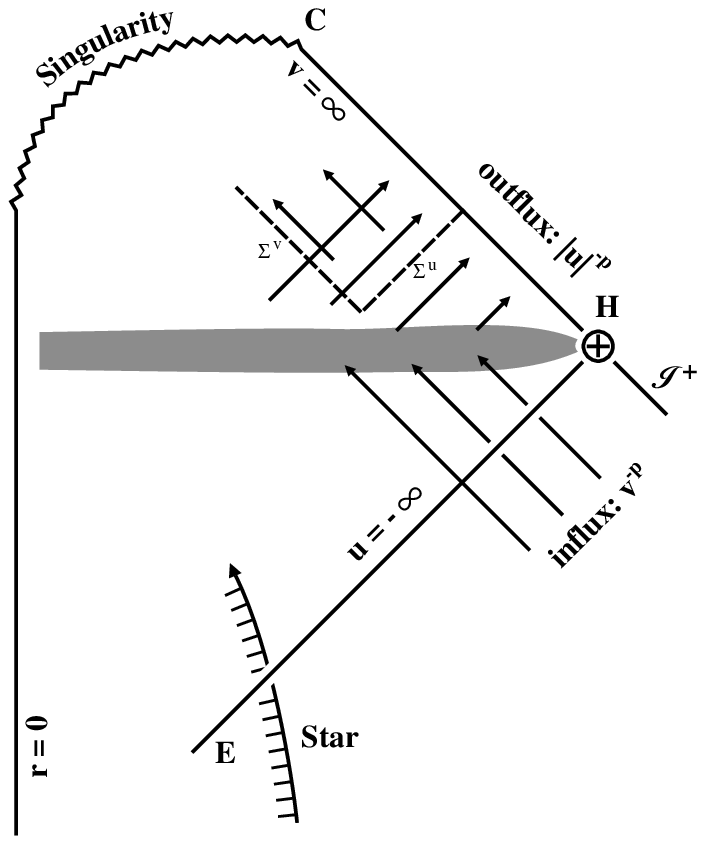}
\begin{quote}
\small
Figure 1.\ The Penrose diagram of a stellar collapse leading to a
black hole. The region above the potential barrier (gray) is
modeled to a high accuracy by a plane wave space time. The dashed
lines represent the characteristic initial surfaces used to
numerically solve Einstein's equations. Note that the ``point'' H is
not part of the spacetime, but is an artifact of the mapping.
\end{quote}

Second the high blueshift of the infalling radiation only occurs
behind the interior gravitational potential which scatters this
radiation once
again \cite{bonnano:95}. Thus infalling radiation will propagate on a
nearly stationary background. Only after the scattering occurs will
the blueshift set in at which point the back reaction onto the
spacetime can no longer be neglected. This leaves us effectively with
a cross flow
situation near the Cauchy horizon. Figure 1 summarizes the
situation.

Let us now focus our attention to this inner region near the Cauchy
horizon.
The most general metric we can write down is of the form
\cite{brady:96}
\begin{equation}
    ds^2 = -2 e^\lambda du dv
                + g_{ab} \left(d\theta^a + s^a du\right)
                 \left(d\theta^b + s^b du\right)
    \label{eq:generalds}
\end{equation}
where the lower case indices run from $2$ to $3$ and $\theta^a$
denote spacelike
coordinates. A more detailed discussion of a metric of the form
(\ref{eq:generalds}) appears in \cite{brady:96}.

To get some idea of  the expected behaviour of the metric
coefficients in
(\ref{eq:generalds}) let us draw the comparison to the spherical case
described in more detail in \cite{bonnano:95}. For a spherical line
element we have $s^a = 0$ and  $g_{ab} d\theta^a d\theta^b = r^2
d\Omega^2$ in the usual polar coordinates. We therefore would expect
$g_{ab}$ to stay finite as the
Cauchy horizon is approached. The important point to note is that $r$
stays finite at the Cauchy horizon. Thus we do not deal with a $r=0$
type singularity of the type expected deeper in the hole.

The gravitational perturbations of the Reis\-sner-Nord\-str\"om
spacetime which, because of sphericity, must be
modeled by a scalar field, decay as an inverse power law in the
advanced and retarded
coordinates $u$ and $v$ \cite{price:72}. In these coordinates we have
$\exp(-\lambda) \sim \exp \kappa_0 (u+v)$.
Raychauduri's equation then
dictates
\begin{equation}
  r_{,v} \sim v^{-q} \ {\rm and }\  r_{,u} \sim v^{-p},
  \label{eq:r}
\end{equation}
so that
\begin{eqnarray*}
 \lefteqn{- 2  r_{,v} r_{,u} e^{-\lambda}  =  g^{\mu\nu} r_{,\mu}
r_{\nu} } \\
& & = 1 - \frac{2 m(u,v)}{r} + \frac{Q^2}{r^2} \sim e^{\kappa_0
(v+u)}
 v^{-q} u^{-p}/r_-
\end{eqnarray*}
as $v \rightarrow \infty$. Here
$\kappa_0$
is the surface gravity of the unperturbed inner horizon and $p=q-2$
is
related to the multipole moment $l$ of the infalling perturbation by
$p=4l + 4$.

This divergence signifies a scalar curvature singularity, which is
seen for example by inspecting the Kretsch\-mann invariant
$R_{\alpha\beta\gamma\delta}R^{\alpha\beta\gamma\delta} \sim
e^{-2\lambda}\sim \exp(2
\kappa_0 (v+u))$ as $v \rightarrow \infty$.

Assuming that a similar behaviour occurs in the ge\-ne\-ric case,
described by
the metric (\ref{eq:generalds}), we can expand Einstein's vacuum
equations
$R_{\alpha\beta}=0$ in powers of $\exp(-\lambda)$. We can then try to
solve
the zeroth order equations and hope that this solution indeed
satisfies the assumptions made in the expansion.

Performing such an expansion for the metric  (\ref{eq:generalds})
yields equations in $u$ and $v$ only for the metric functions
\cite{brady:96:2}, i.e.\ all $\theta^a$-derivatives are multiplied by
at least a factor of $\exp(\lambda)$. Furthermore it turns out that
for Price law initial
conditions the functions $s^a \sim \exp(\lambda)$ \cite{brady:96:2}
so
that it effectively drops out of the zeroth order equations.

Thus to a very good approximation the vicinity of the Cauchy horizon
is described by the metric
\begin{equation}
 ds^2 = -2 e^{-\lambda}du dv + g_{ab} d\theta^a d\theta^b.
 \label{eq:ds}
\end{equation}
Now the functions $\lambda$ and $g_{ab}$ only depend on the
coordinates $u$ and $v$. The metric (\ref{eq:ds}) describes the
interaction region of two plane gravitational waves. This is not an
accident, but has a deeper physical meaning.  Remember, that the
Cauchy horizon lies at a finite radius $r=r_-$. Thus on length
scales $l < r_-$ the inner horizons looks ``flat'' and the
gravitational waves can be approximated by plane waves.
Mathematically this corresponds to replacing $r^2 d\Omega^2$ by
$g_{ab}(u,v)d\theta^a d\theta^b$. This fact is also reflected in the
way the curvature diverges at the Cauchy horizon. Rather than $1/r
\rightarrow \infty$ we have have the metric function $\exp(-\lambda)
\rightarrow
\infty$ as $v \rightarrow \infty$.
But the metric (\ref{eq:ds}) is only a valid approximation if the
two-metric
$g_{ab}$ stays regular (in a sense defined below). Thus for the
analysis of
the full problem (described by the metric (\ref{eq:generalds})) to be
self-consistent the simple plane wave model must show the mass
inflation behaviour.

We are therefore left with a two dimensional characteristic initial
value
problem
in $u$ and $v$ with initial data placed on the characteristics
$\Sigma^v: u=u_0$ and $\Sigma^u: v=v_0$ (see Figure 1).

Einstein's equations for the metric (\ref{eq:ds}) are
\cite{brady:96}:
\begin{eqnarray}
    \lefteqn{0  =  e^{\lambda}\; R_{ab} = } \label{eq:Rab}\\
    & 4 \sigma_{vd(a} \sigma_u{}^d_{b)} - 2 \partial_{(u}
\sigma_{v)ab}
    + g_{ab} \left(K_u K_v + \partial_{(u} K_{v)} \right) \nonumber
\\
    \lefteqn{0  =  R_{uv}  = \partial_{(u} K_{v)} -\lambda_{,uv} -
     \sigma_{uab} \sigma_v{}^{ab} - \frac{1}{2} K_u K_v }
    \label{eq:Ruv}
\end{eqnarray}
and
\begin{eqnarray}
  0 = R_{vv} & = & K_v \partial_v \lambda - \sigma_{vab}
\sigma_v{}^{ab} -
     \frac{1}{2} K_v^2 - \partial_v K_v,
  \label{eq:Rvv} \\
  0 = R_{uu} & = & K_u \partial_u \lambda - \sigma_{uab}
\sigma_u{}^{ab} -
     \frac{1}{2} K_u^2 - \partial_u K_u
  \label{eq:Ruu}
\end{eqnarray}
where lower case
indices are raised and lowered with the two-metric $g_{ab}$. The
trivial equation (\ref{eq:Ruv})  follows from  equations
(\ref{eq:Rvv}) or  (\ref{eq:Ruu})  by simple differentiation.

The quantity
\begin{equation}
    \sigma_{Aab} = \frac{1}{2} \sqrt{g} \
       \partial_A \left(\frac{g_{ab}}{\sqrt{g}}\right)
    \label{eq:shear}
\end{equation}
 is the shear of the congruence
$\ell^{(A)}_\alpha dx^\alpha =
du^A$,
\begin{equation}
    K_A = \partial_A \sqrt{g}
    \label{eq:expansion}
\end{equation}
is its expansion and we have defined  $g := det(g_{ab})$. These
quantities  are again discussed in more
detail
in \cite{brady:96}.

Taking the trace $g^{ab} R_{ab}$ of equation (\ref{eq:Rab}) gives
$\partial_A K^A + K_A K^A = 0$ or, using the definition
(\ref{eq:expansion}) for $K_A$,
\begin{equation}
    \partial_u \partial_v \sqrt{g} = 0
    \label{eq:harmonicg}
\end{equation}
so that
\begin{equation}
    \rho :=  \sqrt{g} = \rho_0 + F(v) + G(u).
    \label{eq:rho}
\end{equation}

Comparing this expression to the behaviour of $r$ in the spherical
case (\ref{eq:r}) we set
\begin{equation}
        \rho  = \rho_0 +v^{-q+1} - |-u|^{-p+1}.
    \label{eq:rhouv}
\end{equation}
Note that other choices of the functions $F$ and $G$ merely
correspond
to different gauge choices of the null coordinates $u$ and $v$.

Once this choice has been made one can specify the boundary data
for the shear $\sigma_A$ on the characteristics $\Sigma^A$ which
comprises the free gravitational data.

The traceless part of equation (\ref{eq:Rab}) is
\begin{equation}
    \partial_{(u} \sigma_{v)ab} - 2 \sigma_{ud(a}
\sigma_v{}^{d}{}_{b)} =
0.
    \label{eq:shearwave}
\end{equation}

Once we have solved this equation for $\sigma_{Aab}$ we can use the
constraint equation (\ref{eq:Rvv}) or (\ref{eq:Ruu}) to solve for
$\lambda$.

To be more specific let us choose a definite form for the two-metric
$g_{ab}$. A convenient parameterization is \cite{yurtsever:89}
\[
 g_{ab} = \rho \left[
   \begin{array}{cc}
    e^{2\beta} \cosh(\gamma) & \sinh(\gamma)  \\
    \sinh(\gamma) & e^{-2\beta} \cosh(\gamma)
   \end{array}
   \right].
\]

The matrix equation (\ref{eq:shearwave}) then is equivalent to the
two
coupled nonlinear PDE's
\begin{equation}
        \beta_{,uv} + \frac{1}{2\rho} \left(\beta_{,u} \rho_{,v}+
        \beta_{,v} \rho_{,u} \right) =
         - \tanh(\gamma)
\left(\beta_{,u}\gamma_{,v}+\beta_{,v}\gamma_{,u} \right)
    \label{eq:gabe}
\end{equation}
\[
        \gamma_{,uv} + \frac{1}{2\rho} \left(\gamma_{,u} \rho_{,v}+
        \gamma_{,v} \rho_{,u} \right) = 2 \sinh(2 \gamma)
        \beta_{,u}\beta_{,v}.
\]
The constraint equations (\ref{eq:Rvv}) and (\ref{eq:Ruu}) are
ordinary first order differential
equations
\begin{eqnarray*}
    \lefteqn{\lambda_{,A} =} \\
     &\frac{1}{\rho_{,A}} \left(\rho_{,AA} -
    \frac{(\rho_{,A})^2}{2\rho} + 2 \rho \cosh^2\gamma (\beta_{,A})^2
+
    \frac{\rho}{2}(\gamma_{,A})^2 \right)
\end{eqnarray*}
for $A = u,v$.
It is convenient to define a new function $\Lambda := \lambda + 1/2
\ln \rho - \rho_{,v} - \rho_{,u}$ in terms of which the above
equations
become
\begin{equation}
    \Lambda_{,A} =
     \frac{1}{\rho_{,A}} \left( 2  \cosh^2\gamma (\beta_{,A})^2 +
    \frac{1}{2}(\gamma_{,A})^2 \right)
    \label{eq:sigmaA}
\end{equation}
Note that for our choice of $\rho$ the right hand side of equation
(\ref{eq:sigmaA}) is always smaller or equal to zero for $A=v$.
 Thus we already can conclude that the
function
$-\Lambda$ grows without bound in the limit $v
\rightarrow \infty$
as expected from the spherical case.

Let us now discuss the boundary data for $\beta$ and $\gamma$ which
carry the gravitational degrees of freedom.
If we ignore the nonlinear terms in the wave equations for $\beta$
and $\gamma$, equations (\ref{eq:gabe}) can be rewritten as
\[
  \Box \beta = \Box \gamma = 0
\]
where $\Box$ is the wave operator corresponding to the metric
(\ref{eq:ds}).  Thus Price's analysis
\cite{price:72} tells us that the initial data  on $\Sigma^v$ is
\[
  \beta_{,u} \sim \gamma_{,u} \sim |-u|^{-p/2}
\]
and on
$\Sigma^u$ is
\[
  \beta_{,v} \sim \gamma_{,v} \sim v^{-q/2}
\]
where as before $q=p+2$.

Inserting this into equations (\ref{eq:sigmaA}) and solving for an
asymptotic series indeed gives, at least on the initial surfaces
\[
  \Lambda_{,v} = -\kappa_v^2 + O(1/v) \ {\rm and}\
  \Lambda_{,u} = -\kappa_u^2 + O(1/u)
\]
where  $\kappa_v^2$ and $\kappa_u^2$ are constants. This is the
expected behaviour for $\Lambda$.

This behaviour was indeed found before by Brady and Chambers
in \cite{brady:95} where Einstein's equations for the full metric
(\ref{eq:generalds}) were integrated on a pair of intersecting null
surfaces.

The important question however is if this behaviour still occurs away
from the initial surfaces. In our simple model this is solely
determined by the behaviour of the functions
$\beta$ and $\gamma$, away from the initial surfaces.

In reference \cite{brady:96:2} perturbative arguments are given which
show that for Price type initial data the shear must decay like a
Price tail all
the way down to the inner horizon.
The proof however works only for Price power law initial data. It
would be desirable to see how stable the mass inflation scenario is
towards changes in
the initial data.  A numerical integration of equations
(\ref{eq:gabe}) and (\ref{eq:sigmaA}) gives the possibility to study
the behaviour of the functions $\beta$ and $\gamma$ in detail for
different types of boundary data, e.g Price power law or a Gaussian
pulse. It furthermore provides a direct verification of the
pertubative agruments presented in reference \cite{brady:96:2}.


\section{Numerical solution of the shear equations}

The most natural and obvious approach to the problem is to
numerically integrate the characteristic initial value problem for
equations (\ref{eq:gabe}) and then use the obtained values for
the functions $\beta$ and $\gamma$ to calculate $\Lambda$ using
(\ref{eq:sigmaA}). While this approach has some problems (see below)
it is nevertheless the most straightforward way to numerically
integrate these
equations. The numerical procedure, which is schematically displayed
in the following flow chart, is second order in $v$ and first order
in $u$.

During our numerical experiments we have used Price power law initial
data for $\beta$ and a power law or Gaussian pulse for $\gamma$ on
the characteristic initial surfaces $\Sigma^u$ and $\Sigma^v$. We
performed runs for various different values of the parameters $p$ and
$q$ as well as different amplitudes for the initial tails.  While
a Gaussian pulse as initial data changes the simple behaviour of
$\Lambda \sim \exp(-\kappa_v^2 v - \kappa_u^2 u)$  it still leads to
the
qualitatively same result, namely a divergence of $\Lambda$ as
described above. Power law initial data for both functions $\beta$
and $\gamma$ on the other hand show the behaviour expected from the
analytical arguments.
\mbox{
    \begin{picture}(220,350)(-10,-10)
       \put(90,337){\vector(0,-1){20}}
       \put(10,290){\fbox{\parbox{160pt}%
       {Calculate $\beta$ and $\gamma$ on $\Sigma^{u_i}$ from
        $\beta_{,v}$ and $\gamma_{,v}$ using a second order
method.}}}
        \put(90,271){\vector(0,-1){20}}
        \put(10,224){\fbox{\parbox{160pt}%
       {From this calculate $\Lambda$ on $\Sigma^{u_i}$ using the
$R_{vv}=0$
       equation.}}}
       \put(90,205){\vector(0,-1){20}}
       \put(10,158){\fbox{\parbox{160pt}%
       {Insert this into $R_{uu}$ to determine the numerical
accuracy.}}}
       \put(90,139){\vector(0,-1){20}}
       \put(10, 92){\fbox{\parbox{160pt}%
       {Use the wave equations (\ref{eq:gabe}) to calculate
        $(\beta_{v})_{,u}$ and $(\gamma_{v})_{,u}$.}}}
       \put(90,73){\vector(0,-1){20}}
       \put(10, 26){\fbox{\parbox{160pt}%
       {Calculate $\beta_{,v}$ and $\gamma_{,v}$ on
$\Sigma^{u_i+\delta
       u}=\Sigma^{u_{i+1}}$ using a first order method.}}}
       \put(90,7){\vector(0,-1){10}}
       \put(90,-3){\vector(-1,0){45}}
       \put(45,-3){\line(-1,0){45}}
       \put(0,-3){\vector(0,1){165}}
       \put(0,162){\line(0,1){166}}
       \put(0,328){\vector(1,0){45}}
       \put(45,328){\line(1,0){45}}
    \end{picture}}
\begin{quote}
 \small
  This flow chart shows the principal workings of the program used to
solve equations (\ref{eq:gabe}). $\Sigma^{u_i}$ denotes the surface
of constant
  $u=u_0 + i \delta u$, where $\delta u$ is the ``time'' step of the
numerical grid.
\end{quote}
In  Figure 2 below we show $\Lambda$ for such an
experiment.

The problem with the characteristic initial value formulation is the
assumption of an already existing portion of the Cauchy horizon. We
thus excluded by hand the possibility of a spacelike singularity
occurring before a inner horizon could form. This is acceptable
insofar as one is looking for a self-consistent description of the
structure of the Cauchy horizon. If there were a spacelike
singularity the approximation leading from the general metric
(\ref{eq:generalds}) to the plane wave metric (\ref{eq:ds}) would be
invalid. On the other hand it would be satisfying to know that
equations (\ref{eq:gabe}) do not lead to a spacelike singularity for
generic Price power law initial conditions, placed on a spacelike
initial surface.

To numerically check this one has to change from the null coordinates
$u$ and $v$ to a set of space- and timelike coordinates.
This is actually quite easy in the present situation thanks to
equation (\ref{eq:harmonicg}) for $\rho$. This equation is the
integrability condition for the coordinate transformation
\begin{equation}
    \begin{array}{rr}
	\rho  = & \rho_0 + F(v) + G(u)  \\
	\chi  = &  F(v) - G(u)
     \end{array}
    \label{eq:trans}
\end{equation}
\renewcommand{\arraystretch}{1.0}%
where $F$ and $G$ have been specified in  (\ref{eq:rhouv}). This
transforms the line element (\ref{eq:ds}) into
\begin{equation}
	ds^2 = \frac{e^\Lambda}{\sqrt{\rho}} \left(d\chi^2 - d\rho^2 \right)
+
	g_{ab} d\theta^a d\theta^b.
	\label{eq:newds}
\end{equation}
Note that this coordinate transformation only works because
$\rho = \sqrt{g}$ is harmonic. In other words we can only perform
this transformation near the Cauchy horizon, where the metric
(\ref{eq:ds}) is valid. Thus, even though the transformation is
defined for $v>0$ and $-u > 0$ it only is applicable near the Cauchy
horizon, i.e. in the limit of large $v$.

%
%

\epsfxsize=8.0cm
\epsffile{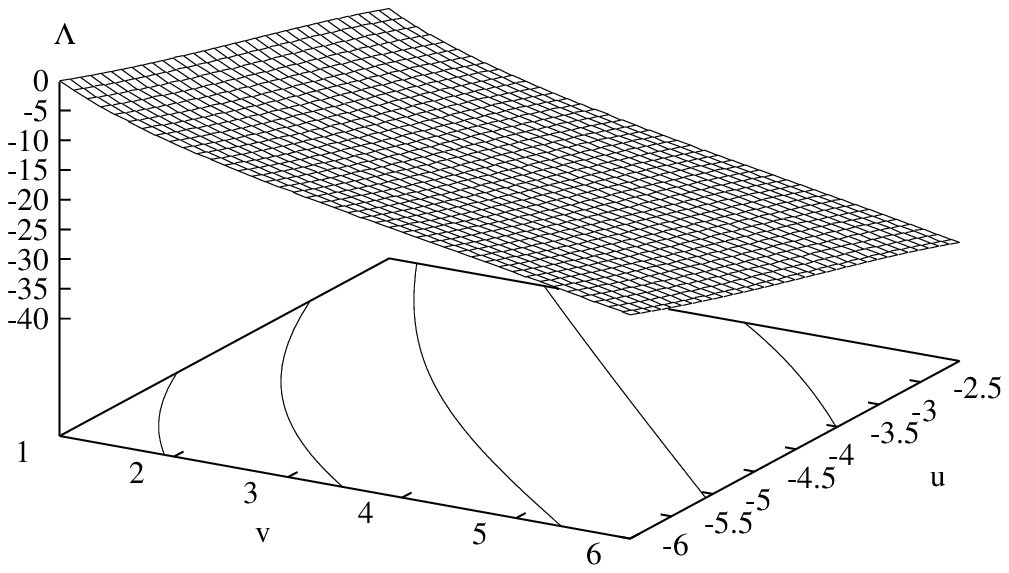}
\begin{quote}
\small
Figure 2.\ Shown is the function $\Lambda$ which corresponds to the
negative
   logarithm of the ``mass function'' of the hole. Thus $m$ increases
   exponentially towards the inner horizon ($v \rightarrow \infty$).
\end{quote}

Initial data is now placed on the surface $\rho=\rho_0$. (Note that
$\rho$, which in the spherical case is proportional to the radial
coordinate $r^2$ is timelike, see Figure 3.)  Only this surface goes
into the
``corner''
H of Figure 1. For $\rho<\rho_0$ it intersects the inner horizon
while,
for  $\rho>\rho_0$ it seems to intersect the event horizon. However
the plane wave metric (\ref{eq:ds}) is not a valid approximation
of the spacetime in the limit $u \rightarrow -\infty$, $v=constant$.

The equations governing the functions $\gamma$ and $\beta$ in terms
of the new coordinates are
\begin{eqnarray*}
	\beta_{,\rho\rho}-\beta_{,\chi\chi}+\frac{\beta_{,\rho}}{\rho} & = &
	 2 \tanh(\gamma) \left( \beta_{,\chi} \gamma_{,\chi} -
\beta_{,\rho}
	 \gamma_{,\rho} \right) \\
	 \gamma_{,\rho\rho}-\gamma_{,\chi\chi}+\frac{\gamma_{,\rho}}{\rho} &
= &
	   2 \sinh(2 \gamma) \left( \left(\beta_{,\rho}\right)^2 -
	   \left(\beta_{,\chi}\right)^2\right).
\end{eqnarray*}

The two constraint equations for $\Lambda$ now become
\renewcommand{\arraystretch}{1.8}
\begin{equation}
    \begin{array}{rr}
	\Lambda_{,\rho}  = & 2 \rho \cosh^2(\gamma)
	\left( \left(\beta_{,\rho}\right)^2 +
	   \left(\beta_{,\chi}\right)^2\right) \\
	 &   \mbox{} + \frac{\rho}{2} \left( \left(\gamma_{,\rho}\right)^2 +
	   \left(\gamma_{,\chi}\right)^2\right)
    \end{array}
    \label{eq:newbeta}
\end{equation}
\renewcommand{\arraystretch}{1.0}%
and
\[
	\Lambda_{,\chi}  =  4 \rho \cosh^2(\gamma) \beta_{,\chi}
	\beta_{,\rho} + \rho \gamma_{,\chi} \gamma_{,\rho}.
	\label{eq:newgamma}
\]

%
%
\begin{center}
\epsfxsize=8.0cm
\epsffile{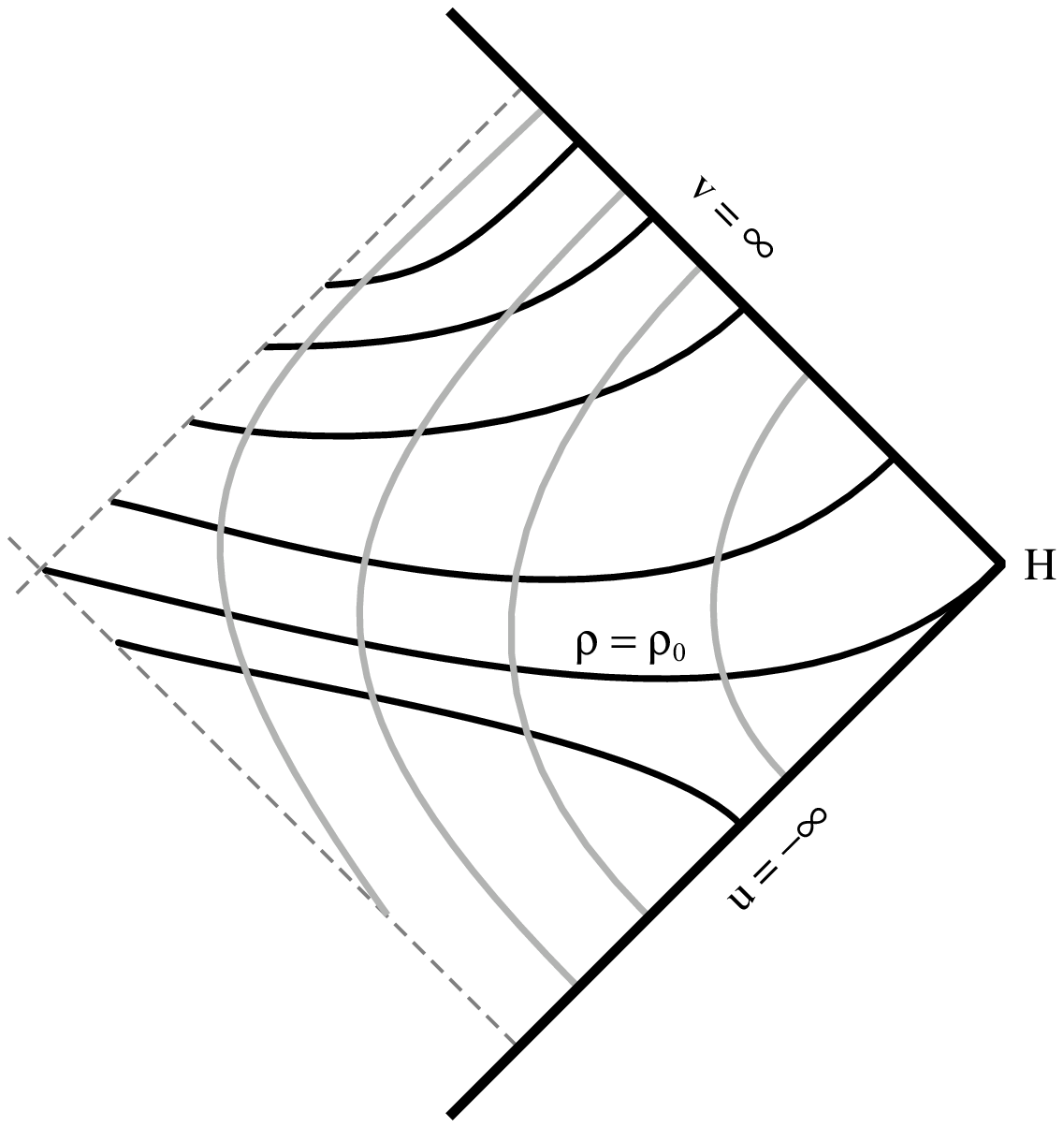}
\begin{quote}
\small
Figure 3.\ This Figure shows the new coordinates $\rho$ and $\chi$.
Initial data is now placed on the surface $\rho=\rho_0$, and the
evolve to smaller values of $\rho$, i.e. towards the inner horizon.
\end{quote}
\end{center}

Again, because the right hand side of equation (\ref{eq:newbeta}) is
always greater or equal to zero $\Lambda$ has to decrease as we go to
smaller values of $\rho$, i.e. towards the Cauchy horizon. While we
still expect the functions $\beta$ and $\gamma$ to stay bounded this
is not true any more for their derivatives, since terms like
$\partial
v/
\partial \rho$ which come into the coordinate transformation blow up.
Thus we still expect $\Lambda$ to diverge as we approach the inner
horizon.

To integrate these equations we have used a centered two step
Lax-Wendrof scheme \cite{press:95} which is second order in $\rho$
and $\chi$. The program now evolves the initial data  inside the
domain of dependance of the initial Cauchy surface, so that no
assumptions
about boundary data have to be made. For appropriate values of
the numerical grid constants this domain covers almost the entire
region between the surface $\rho = \rho_0$ and the inner horizon.

Again the results agree with the analytical approximations as well as
the characteristic numerical results indicating the validity of the
characteristic code. In particular we don't see
any spacelike singularity developing. It must be added however that
even this approach is not completely useful in determining this
question. While the initial surface $\rho=\rho_0$ approaches the
``corner'' H it also  approaches the inner horizon, which initially
is
at radius $\rho=\rho_0$.  Ideally one would like to deal with a space
like
Cauchy surface that never touches the inner horizon, but still always
stays below the potential barriere. However this cannot be
accomplished by the
simple coordinate transformation (\ref{eq:trans}). On the other hand
the two surfaces only meet at infinitly late time $u=-\infty$ and
$v=\infty$.

\section{Conclusion}

The main purpose of this paper is to support and supplement the
analytical work done \cite{poisson:89,bonanno:95,brady:96:2} on the
structure of the Cauchy horizon or mass inflation singularity. While
the latter is necessarily confined to perturbative methods a
numerical
analysis gives one the opportunity to investigate specific examples
for nontrivial realistic initial data.

The results of our analysis strongly support the result of the
analytical work. We indeed find the predicted behaviour of the shear
and the metric function $\Lambda$, which is essentially the negative
logarithm of the (quasi-local \cite{israel:96}) mass function of the
black hole. It thus seems that mass inflation singularity
is a generic singularity \cite{brady:96:2,ori:95}.

\end{document}